# A Comparison of Online Automatic Speech Recognition Systems and the Nonverbal Responses to Unintelligible Speech

Joshua Y. Kim[1], Chunfeng Liu[1], Rafael A. Calvo[1]*, Kathryn McCabe[2],
Silas C. R. Taylor[3], Björn W. Schuller[4], Kaihang Wu[1]

[1] University of Sydney, Faculty of Engineering and Information Technologies
[2] University of California, Davis, Psychiatry and Behavioral Sciences
[3] University of New South Wales, Faculty of Medicine
[4] Imperial College London, Department of Computing

## Abstract

Automatic Speech Recognition (ASR) systems have proliferated over the recent years to the point that free platforms such as YouTube now provide speech recognition services. Given the wide selection of ASR systems, we contribute to the field of automatic speech recognition by comparing the relative performance of two sets of manual transcriptions and five sets of automatic transcriptions (Google Cloud, IBM Watson, Microsoft Azure, Trint, and YouTube) to help researchers to select accurate transcription services. In addition, we identify nonverbal behaviors that are associated with unintelligible speech, as indicated by high word error rates. We show that manual transcriptions remain superior to current automatic transcriptions. Amongst the automatic transcription services, YouTube offers the most accurate transcription service. For non-verbal behavioral involvement, we provide evidence that the variability of smile intensities from the listener is high (low) when the speaker is clear (unintelligible). These findings are derived from videoconferencing interactions between student doctors and simulated patients; therefore, we contribute towards both the ASR literature and the healthcare communication skills teaching community.

## 1 Introduction

ASR systems are continually improving. In recent years, the improved performance of ASR systems has made it possible for them to be deployed in large-scale commercial products such as Google Home and Amazon Alexa. Mainstream ASR systems use only voice as inputs, but there is potential benefit in using multi-modal data in order to improve accuracy [1]. Compared to machines, humans are highly skilled in utilizing such unstructured multi-modal information. For example, a human speaker is attuned to nonverbal behavior signals and actively looks for these non-verbal 'hints' that a listener understands the speech content, and if not, they adjust their speech accordingly. Therefore, understanding nonverbal responses to unintelligible speech can both improve future ASR systems to mark uncertain transcriptions, and also help to provide feedback so that the speaker can improve his or her verbal communication.

With the recent advancements in artificial intelligence, there is a wide range of ASR systems that can produce high-quality transcripts. In this paper, we aim to provide empirical evidence on the performance of five ASR providers - namely, Google Cloud, IBM Watson, Microsoft Azure, Trint, YouTube. We investigated whether ASR services produce transcriptions that are of equivalent quality to the significantly more expensive manual transcription services.

ASR system error rates could potentially result from a variety of causes apart from speech intelligibility. Firstly, they could arise due to recording issues where the conversation participants did not experience any issues during the conversation, but the recording is unreliable because of technical issues like unstable internet connection with the server performing the recording. Secondly, the downstream data processing that converts the recording to the desired file format of the ASR may result in the reduction of audio quality. Thirdly, because the ASR models are trained independently with different training datasets and model architectures, the performance of the ASR models differ. If the performance of an ASR model is poor, it will produce a high word error rate even when given a recording of a clearly articulated speech.

In all these cases, since the communication between the two parties is clear and the poor-quality of the transcription is due to reasons outside of the conversation, the listener would not have displayed nonverbal behaviors that gave hints that he/she could not understand the speaker; therefore, any analysis attempting to quantify the relationship between the nonverbal behavior of the listener and the intelligibility of the speech would be weak. In this study however, we are not concerned with such issues. Instead, we are concerned with those issues that impact speech intelligibility and are experienced by both the listener and the manual or automatic transcriber. For example, noisy recording environments or poor speaker articulation.

Other factors contribute to speech intelligibility. Individual factors may be due to physical or mental illnesses that result in unclear speech, for example, Parkinson's Disease [2]. Culture and ethnicity may also interfere with the intelligibility of speech and the listener's ease of understanding [3]. When the listener experiences speech intelligibility problems they are likely to express this difficulty with both verbal and non-verbal cues. For instance, they may send non-verbal cues that they are not understanding what is being said through their facial expressions [4] and hence poor speech intelligibility may cause certain facial expressions to occur. It is this relationship that is the secondary focus of this paper.

Understanding the relationship between speech intelligibility and facial expressions could then be used by ASR systems to augment the automated decision whether to replace a low-confidence transcribed word into a flag like "[uncertain transcription]". Also, such insights could facilitate downstream research that seeks to improve human-human communication skills by highlighting the range of facial expressions, displayed by the listener, which indicate uncertainty. While our domain is limited to physician communication as our dataset consists of clinical consultations between actor patients and student doctors, we posit that the findings are valuable for both the ASR community and more broadly, to the healthcare communication skills teaching community.

In this paper, we address two research questions. First, we provide empirical evidence to the question, "Which automatic transcription is currently the most accurate?". Second, we investigate the research question, "what is/are the nonverbal behavior(s) of listeners that are associated with high transcription error rates (indicating intelligibility of speech)?"

## 2 Related Works

### 2.1 ASR Word Error Rate (WER) Performance

With the recent advancement brought about by neural network architectures, ASR systems have seen significant improvements over the past few years [5]. The Switchboard telephone speech dataset is often used to benchmark the performance of the transcription [6]–[8]. However, researchers may differ in using different subsets for evaluation. The WER performance provided by each of the vendors is discussed in turn. Microsoft Research reports a WER of 5.1% on the NIST 2000 Switchboard task [5]. IBM Research reports 6.6% WER on the Switchboard subset of the Hub5 2000 evaluation test set [7]. Google Research reports a 6.7% WER on a 12,500-hour voice search dataset and 4.1% on a dictation task [9], both of which are not part of the Switchboard telephone speech dataset. Instead, both datasets are extracted from the Google traffic application, and the two tasks differ in that the dictation task contains longer sentences than the voice search utterances.

Telephone speech or dictation tasks are typically recorded with the microphone located near the speaker. However, applications of the ASR in teleconferences is more challenging as the speaker is speaking at some distance from the microphone – this is known as distant speech recognition [10]. Research on distant speech recognition includes the application of convolutional neural networks (CNN) [11] on the Augmented Multi-party Interaction (AMI) meeting corpus [12], where a word error rate of 40.9% was achieved with a single distant microphone [13]. More recently, Renals and Swietojanski [14] used the AMI corpus to compare ASR approaches using multiple distant microphones and individual headset microphones. The difference in WER is significant - the eight distant microphone setup achieved a WER of 52.0% while the individual headset microphone setup achieved a WER of 29.6%. It is also worth noting that the WER from individual headset microphone setup using the AMI corpus (29.6%) is higher than the WER reported by the vendors using the Switchboard dataset (Microsoft: 5.1%; IBM Watson: 6.6%).

We expect the recordings from our video consultations to be more similar to the performance under distant speech recognition conditions, as the setup is not professionally dedicated to recording clean speech. Këpuska and Bohouta [15] performed a comparison between CMU Sphinx, Microsoft Speech and Google Cloud and found that the Google Cloud API performs the best with a mean WER of 9%. In that study, the authors used the Texas-Instruments/Massachusetts Institute of Technology (TIMIT) corpus [16]. Whilst it is unclear whether the audio is captured from a distance, the low WER suggests that it is not using a distinct microphone setup. In the present study, we expand the number of online transcription services for comparison and utilize a dataset that is intended to mirror real-world doctor-patient interviews. Thus, it is different from previous datasets in that, instead of short utterances, it consists of long professional conversations from real-world scenarios.

## 2.2 Detecting unintelligible speech from the listener's face

In this paper, we investigate whether smiling, head nodding, and frowning is indicative of a confused listener. Since we do not have access to specialized equipment such as facial electromyography to detect nonvisible muscle contractions [17], we instead focus on the literature on detecting nonverbal behavior obtained via analysis of video data.

Referencing the literature on nonverbal behaviors that are associated with cognitive loads, Ekman and Friesen [18] showed that automatic detection and analysis of facial Action Units (AU) is an important building block in the analysis of nonverbal behavior. Smiling (AU12) and frowning (AU04) have been found to be positively associated with self-efficacy in students, who were tasked to listen to a narrative of information while solving a task [19]. The positive relationship between frowning and self-efficacy is, according to the authors, a reflection of mental exertion and not negative affect, such as frustration. The association of frowning with higher cognitive loads is also found in other research studies [20], [21]. Lastly, head nodding is seen as an integral part of backchanneling [22] – a short feedback response such as "uh-huh" [23] – and communicates "message understood" [24]. On the other hand, head shakes may be interpreted as "disapproval" and unfavorable [25].

In studies conducted within our specific domain, i.e. – doctor-patient consultation, Crane and Crane [26] found that the degree of smiling, frowning, and head nodding was predictive of clinical outcomes. This association of nonverbal communication and clinical outcomes is a result of a wide variety of interrelated factors. For example, nonverbal communication has been shown to influence the patient-perceived quality of care [27], improve rapport [28], improve patient understanding of information [29], or improve patient compliance [30]. These factors, in turn, influence better clinical outcomes [31].

For the present study, we analyze the aforementioned gestures (smiling, frowning, head nodding and head shaking). Since our dataset of 24 videos from 12 consultations is relatively small, we use the literature to guide our focus on specific gestures to preserve statistical power.

## 3 EQClinic Dataset

### 3.1 Data collection

This study is an extension of previously collected data from the EQClinic platform [32]. Students in an Australian medical school were required to complete the program aimed at improving clinical communication skills during their first and second years of study. Within the EQClinic platform, the students were required to complete at least one medical consultation with a simulated patient on the online video conferencing platform EQClinic [33]. A Simulated Patient (SP) refers to a human being who has been trained to act as a patient in a medical situation. Briefly, EQClinic works on most web browsers of a PC or an Android tablet, and it uses OpenTok, a Web Real-Time Communication (WebRTC) service, to provide real-time video communication. All consultations conducted on EQClinic are automatically recorded by OpenTok. In this paper, we selected twelve consultations in the year 2016 as the dataset to analyze.

Participants were twelve second-year undergraduate medical students (six female and six male) and two SP (one male and one female). The two SP were professional actors, recruited online and paid $AU35 per hour for participating. The study was approved by the UNSW Human Research Ethics Committee (Project Number HC16048), and all participants completed a signed consent before commencing the study.

### 3.2 Data analysis

**Transcription Process**

For each consultation, EQClinic generated one MP4 video recording for the student and one MP4 video recording for the SP with the resolution of 640x480 pixels, and a frame rate of 25fps. Audio recordings of video consultations were extracted from the video recordings using the FFMpeg software [34].

In our database of videos, there are only two SPs (one male and one female) who were regularly interviewed by student doctors. These two SPs completed a total of 84 interviews in the relevant study period. Of the interview sessions performed by the two regular SP, we selected twelve interview sessions pseudo-randomly as we ensured that there are three videos for each of the possible gender pairing (male-male, male-female, female-male, and female-female). Equal representation of gender pairing ensures that we controlled for gender before performing subsequent correlation analysis between WER and non-verbal behavior measures.

The duration of these sessions ranges from 12 to 18 minutes (mean duration (SD) = 14.8 (2.0)). Each session contained two videos, and each of these video pairs had one speaker (the student or the SP). Each video comprised 668 to 1705 words (mean words (SD) = 1187 (316)). In total, 24 videos and a total of 28,480 words were analyzed. Disfluencies like "um" are captured in the transcripts. We sent these 24 videos to seven transcription services - two of which were manual, and the other five were ASR systems. The transcription processes of each of the seven services are described in the next few paragraphs. The costs and file formats required for transcription are summarized in Table 1 in the supplementary material.

| Service | File Format | Cost (USD per video minute) |
|---|---|---|
| Manual (CB) | MP4 Video | 1.920 |
| Manual (Rev) | MP4 Video | 1.500 |
| Automatic (Google Cloud) | Mono-channel FLAC audio | 0.048 |
| Automatic (IBM Watson) | Mono-channel FLAC audio | 0.020 |
| Automatic (Microsoft Azure) | Mono-channel WAV audio (16,000 samples per second) | 0.008 |
| Automatic (Trint) | MP4 Video | 0.250 |
| Automatic (YouTube) | MP4 Video | 0.000 |

Table 1 – Summary of required file formats and costs for transcription services. CB denotes the independent professional transcriber. Rev denotes transcribers from Rev.com.

For the two manual transcription services, one was an independent professional transcriber (CB), and the other was from an online network of hand-picked freelancers available at Rev.com (Rev). For both manual transcription services, video files were provided in the MP4 format for transcription.

Each of the five ASR services (Google Cloud, IBM Watson, Microsoft Azure, Trint, and YouTube) required a different format of the input file to perform the transcription and FFMpeg was used to do all the necessary file format conversions. We discuss these differences in detail below. Also, for all of the five ASR services, we elected to perform asynchronous transcription service calls so that we could compare the results because YouTube and Trint do not offer synchronous transcription service calls. Synchronous service calls refer to the ability for the ASR to stream text results, immediately returning text as it is recognized from the audio – as opposed to asynchronous service calls, where the text result is only returned after the entire session has been analyzed. Whilst we acknowledge that supplying difference file formats in accordance to the requirements of different service providers meant that the comparison of performance is not strictly comparable, we seek to answer the research question "given the current requirements of the different service providers, which provider gives the most accurate transcription?"

*Google Cloud Speech-To-Text*: Google cloud accepts mono-channel FLAC files as input. FFMpeg was used to perform the conversion from MP4 into the FLAC files. The Python library speech_v1p1beta1 from the Google cloud package was used to submit the FLAC file for transcription. The parameters used in the submission call were model='video', use_enhanced=True and the default language_code 'en-US'. The transcription was generated within 15 minutes.

*IBM Watson Speech-To-Text*: The same FLAC files from the Google Cloud conversion were used for IBM Watson. The Python library SpeechToTextV1 from the watson_developer_cloud package was used to submit the FLAC file for transcription. The parameters used in the submission call were content_type='audio/flac'. The transcription was generated within 15 minutes.

*Microsoft Azure Speech-To-Text*: Microsoft Azure accepts WAV files as input. FFMpeg was used to perform the conversion from MP4 into WAV files with PCM encoding and 16,000 samples per second. We then used the Java library (com.microsoft.cognitiveservices.speech) to submit the WAV files for transcription. In the submission call, we did not use any non-default values. The default values expect a wav file with 16-bit sample, 16kHz sample rate, a single channel (Mono) and region default to 'en-US'. The transcription was generated within 15 minutes.

*Trint:* Trint provides a graphical user interface (GUI) for the user to upload videos and download the transcription. Trint accepts MP4 files, so the same MP4 files supplied to the manual transcribers were uploaded onto the Trint platform. Trint then auto-generated the transcription within 15 minutes.

*YouTube Captions*: Similarly to Trint, YouTube enables users to access the transcription of uploaded videos through a GUI. The same MP4 files were uploaded to the YouTube service and the transcriptions downloaded within a day. It should be noted that the free YouTube service only allows 100 videos per day.

**Post transcription processing**
After the transcripts were collected from each of the seven transcription services metadata such as "inaudible" tags, timestamps and punctuations were removed. All words were set to the lowercase. The result of each post-processing was a set of 24 text files each containing the transcription for one of the two participants in each consultation.

**Computing WER and bootstrapping**
After post-processing was complete, we compared the quality of transcripts gathered from different transcription services. Word Error Rate (WER) is a popular performance measure in automatic speech recognition [35]. It is defined as the edit distance between two transcripts - one being the reference transcript and the other being the hypothesis transcript. The edit distance is defined as the minimum number of insert, substitute, and delete operations necessary to transform one sentence from the hypothesis transcript into the equivalent sentence in the reference transcript [36]. In this study, edit distance was calculated using an open-source library called asr-evaluation [37]. The asr-evaluation library was also used by van Miltenburg et al. [38] to compute the WER for automatic transcriptions obtained from the built-in dictation function from a macOS Sierra 10.12.6.

In this paper, we first determined which of the two sets of manual transcriptions would be the reference transcript. We then compared the five sets of automatic transcriptions against this reference transcript to identify the best performing ASR system. Finally, we performed a correlational analysis between the summary statistics of four visual nonverbal behavior features and transcription WER to investigate the association between nonverbal behavior and the intelligibility of the speech.

To choose which one of the two sets of manual transcription should be the reference transcript and similar to Lippmann et al. [39] and Roy et al. [40] we posit that if multiple transcribers produce similar transcripts as indicated by low WER, they have likely converged on the correct transcription. Therefore, the set of manual transcriptions with the lower WER as compared with each of the five sets of automatic transcription was considered the best choice as the set of reference transcripts. In our analysis, ten pairwise WER from asr-evaluation were generated for each of the five hypothesis transcripts and the two manual sets of transcripts (Manual CB and Manual Rev) consistent with methods reported by Belambert [37].

For the ten pairwise WER estimates, we determined which of the WER-reference pairs were statistically significantly different. To do that, we needed the 95% WER confidence interval. Since the assumptions in classical statistics, e.g. – independent error rates [41] – are not applicable when we fixed the hypothesis transcript to be from one ASR service, we elected to use bootstrapping to generate confidence intervals. The bootstrap technique is used to quantify the uncertainty associated with the WER in our application and involves creating 10,000 bootstrap datasets [42] produced by random sampling with replacement [43]. With the 10,000 bootstrap samples, we computed an average WER. Then, we computed the 95% WER confidence interval by eliminating the top and bottom 2.5% values for the speaker-level WER as well as differences in WER between two services [44].

After establishing the set of manual transcription that was of higher quality, we used this set of manual transcription as our reference transcription to examine the WER of all other transcription services. Next, we established whether differences in WER performance between each transcription service were statistically significant. To do this, we used one set of reference transcription and computed the difference in WER between service X and service Y for each of the 24 transcriptions. Similarly, we then bootstrapped the differences in WER between the two services (service X and Y) and generated the confidence intervals for the differences using 10,000 samples.

**Nonverbal behavior features analysis**

To investigate correlations amongst nonverbal behavior features and WER, we detected four visual nonverbal behavior features for students and SP from each frame of video recordings: smiling, frowning, head nodding and head shaking. We extracted the features using OpenFace 2.0 [45], which is an open source toolkit of facial landmark detection and facial action unit recognition. Using OpenFace, we extracted two selected facial Action Units (AU) based on the Facial Action Coding System (FACS; [18]): AU12 (lip corner puller) and AU04 (brow lowerer). OpenFace measured the intensity of each selected AU in a value range of 0 to 5 (a higher value indicates higher intensity). Then we used the value of AU12 and AU04 as indicators of people's smile [46], [47] and frown [48] intensity respectively. We calculated the mean and standard deviation values of each selected AU feature in line with other researchers that uses Openface for facial analysis [49]–[51]. The head nodding and head shaking gestures were detected by tracking the movement of the nose landmark [52], which also enabled us to identify the start and end times of each gesture. In this paper, the frequency (number of gestures per minute) of nodding and shaking were the extracted features. In total, we extracted six measurements – two relating to the mean of the smile and frown; two relating to the variability of the smile and frown; and two relating to the frequency of head nodding and shaking in the session. Head nodding and shaking frequency are count statistics from the whole session, and therefore not suitable for standard deviation measures. These six measurements are chosen because of the literature discussed in the earlier section 2.2.

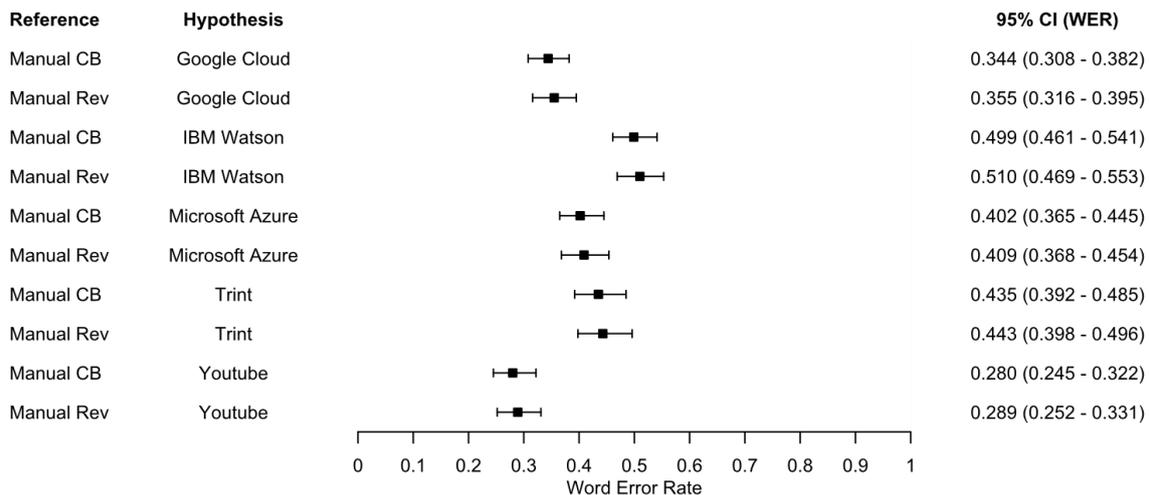

Figure 1 – Forest plot of WER of automatic transcription services, using two sets of reference transcripts from each of the two manual transcription services (Manual CB and Manual Rev). The 95% confidence interval is generated using 10,000 bootstrap samples.

## 4 Results

### 4.1 WER of manual and automatic transcriptions

Figure 1 compares the hypothesis transcripts and each of the two manual transcripts (Manual CB and Rev). The confidence intervals are generated using 10,000 bootstrap samples as described in section 3.2.3. We found that the two sources of manual transcription did not differ significantly. For a given set of hypothesis transcript (generated by selected ASR systems), the confidence interval of Manual CB does not differ from Manual Rev.

Having established that the quality of the two manual transcriptions was similar. We selected Manual CB as the reference transcript and completed a pairwise analysis for the remaining transcription services comparing the quality of all of the transcription services. Figure 2 shows the differences in WER between services pairs. Again, for each of the pairwise difference in WER at a video level, we performed bootstrapping to generate 10,000 samples and compute the 95% confidence intervals. If the 95% confidence interval does not intersect 0, then we conclude that the difference in the pair is statistically significant.

Figure 2 shows that the Manual Rev was the best transcription service, exhibiting significantly better performance relative to the other transcription services. In our case, manual transcription was better than all of the automatic transcription services and all pair-wise differences are statistically significant. Of the automatic transcription services, we found that YouTube exhibited significantly better performance relative to the other automatic transcription services, and all pair-wise difference are statistically significant.

### 4.2 Correlation between WER and nonverbal behavior

Table 2 shows the correlational analysis between the listener's extracted nonverbal behavior and the WER of the speakers from the 24 video recordings of students and SP. WER of each video was calculated by comparing two transcripts - transcripts of CB were used as the reference - and the transcript from Manual-Rev service is used as the hypothesis transcript. The average WER is 17.4% and the standard deviation of WER is 6.92%.

There are two main results from Table 2. Firstly, the intelligibility of the speech is negatively correlated with the standard deviation of smile intensity. In other words, the clearer the speech (lower WER), the higher the variability of smile intensity from the listener. Secondly, it is also worth mentioning that the WER has a negative trend-level effect (p-value = 0.08) with the mean of smile intensity. In other words, the clearer the speech (lower WER), the higher the mean smile intensity from the listener. We did not find evidence that the mean (or standard deviation) of frowning nor the frequency of head nodding or shaking to be associated with WER.

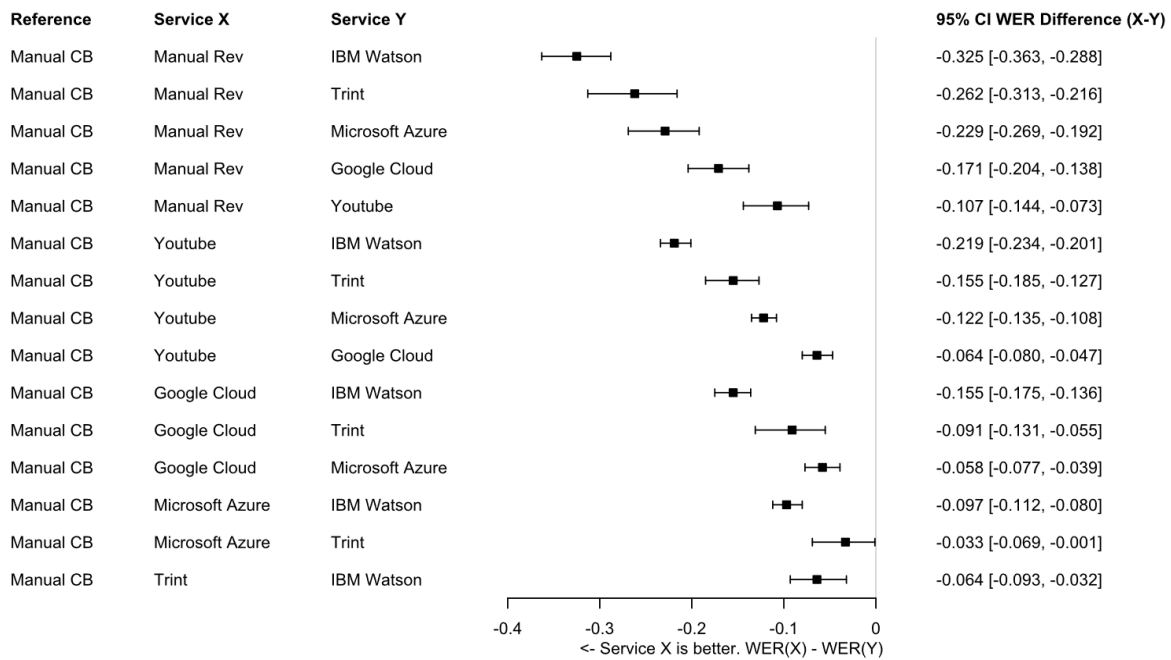

Figure 2 – Forest plot of pair-wise differences in WER of the various transcription services, using Manual CB as the set of reference transcripts. The 95% confidence interval is generated using 10,000 bootstrap samples. For brevity, only comparisons where Service X is better are illustrated. The plot is ordered by the best performing service in Service X, followed by the mean WER difference between Service X and Service Y.

| Feature | Correlation Coefficient | p-value |
|---|---|---|
| Mean of frown intensity | -0.021 | 0.92 |
| Standard deviation of frown intensity | -0.054 | 0.80 |
| Mean of smile intensity | -0.365 | 0.08 |
| Standard deviation of smile intensity | -0.515 | 0.01** |
| Head nodding frequency | -0.128 | 0.55 |
| Head shaking frequency | -0.033 | 0.88 |

Table 2 – Pearson's correlation test between speaker's WER and nonverbal behavior of the listener (N=24). **: < 0.01; *: < 0.05.

## 5 Discussion

In this study, we have two main findings. Firstly, amongst the automatic transcription services, YouTube offers the most accurate automated transcription service, though this is not as accurate as the professional transcription service. Secondly, we found that when the speaker has clear speech, the variability of the listener's smile intensity increases. We discuss these two findings in turn before concluding with a discussion on the limitations and future directions of this study.

### 5.1 Performance of online transcription services

In this study, we used human transcribers and ASR systems to transcribe videoconferencing medical conversations. We found that the two manual transcriptions demonstrated similar quality with WER of 17.4%. This is higher than the WER of previous studies based on the standard telephone audio recording dataset where the manually transcribed WER was between 5.1% and 5.9% [53].

Several potential factors may cause the lower accuracy (that is high WER) of human/manual transcription in this study. First, the conversation environment could have influenced recording quality. The WER in Xiong et al.'s work [53] was tested based on telephone audio recordings, in which the microphone was located near the speaker. However, the medical conversations of this study were conducted over video conferencing on PC or tablets. There was likely to be greater variability in recording quality as some of the speakers were likely seated further away from the microphone. Also, the medical conversation could be held anywhere; therefore environmental noise and audio feedback in the conversation may have impacted the human transcription. Example 1 illustrates a difference between the two manual transcriptions. In this example, the recording was performed with a fan turned on towards the microphone during the recording, hence the recording is filled with background noise. It is assumed that the human transcribers were influenced by environmental noise and audio feedback in this case.

| Manual CB | Manual Rev |
|---|---|
| are there any other symptoms you feel when that happens like any other pain anywhere else soreness sorry could you please repeat that theres a bit of feedback | are there any other symptoms you feel when that happens like any other pain anywhere else soreness sorry im sorry could could you please repeat that theres a bit of feedback |

Example 1: Example transcription of video recorded with a fan turned towards the microphone, creating background noise in the recording. The color scheme of the errors is as follow, Red - Deletion; Orange - Substitution; Green – Insertion

Secondly, the speaker's verbal expressions or intonation could also have caused some of the inconsistency of human transcription. The inconsistency was particularly obvious when the speakers' speech was fast or soft. For example, as shown in Example 2, the quick short utterance of "that's good" is missed by "Manual Rev".

| Manual CB | Manual Rev |
|---|---|
| sorry i just needed to turn the volume up so i could hear you oh thats no good thats good hi okay | sorry i just needed to turn the volume up so i could hear you oh thats not good thats good hi okay |

Example 2: WER could result from a difference in transcriber's interpretation of the audio, and omission of short backchannel utterances. In this example, the transcription on the right transcribed "thats not good" wrongly and omitted the quick backchannel "thats good". The color scheme of the errors are as follow, Red - Deletion; Orange - Substitution; Green – Insertion

Lastly, we posit that the medical nature of the conversations in our study caused the higher WERs from both the manual transcribers and ASR services. This is because even experienced medical transcribers may not be accurate when certain medical terminology is used.

Although in our study, human transcription was not perfect; we found that human accuracy was higher than the tested ASR systems. Of the tested ASR systems, YouTube Captions service achieved the highest accuracy. These results provided us with a preliminary understanding of the transcription qualities of human and ASR systems on video conferencing data. Our results are in line with Këpuska and Bohouta [15] who found that Google Cloud Speech-To-Text outperformed Microsoft Speech Services.

Accuracy may not be the only consideration when we choose a transcription service and other factors, such as processing time and price, may also need to be considered. Regarding processing time, as one would expect, ASR systems are significantly faster than human transcription. The ASR systems took around 15 minutes to process a 15-minute video, and some services allow such transcription jobs to be run in parallel. However, human transcribers took approximately 1 hour to process a 15 min video (with starting and ending timestamps of sentences), and if there was only one transcriber, the transcriptions had to be completed sequentially. Although some companies significantly enhance the efficiency of human transcription by adopting freelancers through the network (100 videos in 24 hours), scalability is still a fundamental limitation for human transcription.

The price structure of the ASR services varied. From Table 1, we see that human transcriber cost 1.5~2 Australian dollars per minute, whereas the prices of ASR systems were less than 0.3 dollars per minute. However, the price was not an independent factor when comparing services. For example, in our tested service, Trint was the most expensive. However, it enables users to access the service through a graphical user interface without any programming effort. This property is especially important for users without any technical background. On the other hand, although Google Cloud, Microsoft Azure, and IBM Watson offered a lower price, to use the APIs, users had to develop programs with different programming languages, e.g. – Python and Java. It may be worth investigating the cost reduction achieved through having the first pass transcription completed by the best ASR then followed by manual transcription, or seek manual transcriptions only for the [inaudible] / [uncertain transcription] tokens returned from the ASR. Another way of increasing cost efficiency is to ensemble multiple ASR systems [42], [54]–[56], using tools like ROVER [57]. Through ensembling multiple ASR systems, a "voting" or rescoring process can reconcile differences across different ASR system outputs, resulting in a composite ASR output that has lower WER than any of the individual systems.

## 5.2 Association of nonverbal behaviour and unintelligible speech

According to Lippmann et al. [39], the intelligibility of human speech could be measured by the WER of transcription. In other words, if the WER of a transcribed speech by two manual transcribers is high, we can understand that the intelligibility of this speech is low. Therefore, as the second contribution of this paper, we investigated if a listener's nonverbal behavior was associated with the speaker's speech intelligibility (indexed by WER). As the results show, we observed correlations between the listener's smile expressions and WER. Specifically, we found that when the speech intelligibility was higher (evidenced by lower WER), the listeners would present a higher standard deviation of smile intensity and higher smile intensity. This result, to some extent, is similar to the observation of a previous study, in which listeners reported less positive emotions while viewing stuttered speech relative to the fluent speech [58].

Finally, the emotional responses could be reflective of their cognitive load during the conversation. In general, listeners' cognitive processing load increases when listening to less intelligible speech. Consequently, we propose that the nonverbal behavior we detected is influenced by this increased cognitive load. This is in line with the findings of Hess et al. [59] where they found that both affective empathic reactions (like facial expression mimicry) and cognitive load contribute towards facial reactions.

## 5.3 Limitations and future work

There are several limitations that should be considered when interpreting the findings. First, our analysis attributes WER between two manual transcribers to speech intelligibility. However, the environment in which the manual transcriber is listening to the consultation is different from that of the participant. For example, the manual transcriber has the option to replay the conversation, and the participant may be in a noisy environment which affects the ability to listen. Second, the evidence from this paper is limited to a highly professional scenario (medical consultation). Whilst we posit that the finding may be generalizable to non-professional settings, or professional settings in a different domain, say customer service or legal consulting, it has yet to be proven. This could be one avenue for future work in this area. Third, the correlation between nonverbal behavior and WER might be affected by demographic factors such as gender and cultural background. These factors should be examined in future studies. Lastly, due to financial considerations arising from the high costs of manual transcription, in this paper, we only selected a small portion of videos from the platform. Because our dataset consists of only 12 interview sessions performed by two SP, any stereotypical attitudes held - consciously or unconsciously - would bias half of the observations as 12 out of 24 videos consists of nonverbal responses from only two SP. We acknowledge that this is a limitation of our study, and in the future, more videos from a wider variety of interviewers will be analyzed to verify the preliminary findings of this paper.

## 6 Conclusion

We have provided the first comparison of the performance of automated transcription services in the domain of dyadic medical teleconsultation. We found that manual transcription significantly outperformed the automatic services, and the automatic transcription of YouTube Captions significantly outperformed the other ASR services. Also, through analyzing the nonverbal behavior features of the listener, we provided evidence that the variability of smile intensity is high (low) when the speech is clear (unintelligible). We posit that these findings could be generalized to other contexts.

## Acknowledgements

The authors thank Hicham Moad S for his help rendered in scripting for the Microsoft Azure API. RAC is partially funded by the Australian Research Council Future Fellowship FT140100824.